\documentclass[aps,prl,superscriptaddress,amsmath,amssymb, reprint,author-numerical]{revtex4-1}

\usepackage{graphicx}
\usepackage{dcolumn}
\usepackage{bm}

\begin{document}


\title{Frequency non-reciprocity of surface spin wave in Permalloy thin films }

\author{O. Gladii}
\affiliation{
Institut de Physique et Chimie des Mat\'{e}riaux de Strasbourg, UMR 7504 CNRS, Universit\'{e} de Strasbourg, 23 rue du Loess, BP 43, 67034 Strasbourg Cedex 2, France
}
\author{M. Haidar}
\affiliation{
Institut de Physique et Chimie des Mat\'{e}riaux de Strasbourg, UMR 7504 CNRS, Universit\'{e} de Strasbourg, 23 rue du Loess, BP 43, 67034 Strasbourg Cedex 2, France
}
\affiliation{Department of Physics, University of Gothenburg, 41296 Gothenburg, Sweden
}
\author{Y. Henry}
\affiliation{
Institut de Physique et Chimie des Mat\'{e}riaux de Strasbourg, UMR 7504 CNRS, Universit\'{e} de Strasbourg, 23 rue du Loess, BP 43, 67034 Strasbourg Cedex 2, France
}
\author{M. Kostylev}
\affiliation{
School of Physics, M013, University of Western Australia, Crawley, 6009, Western Australia, Australia
}
 \author{M. Bailleul}

\affiliation{
Institut de Physique et Chimie des Mat\'{e}riaux de Strasbourg, UMR 7504 CNRS, Universit\'{e} de Strasbourg, 23 rue du Loess, BP 43, 67034 Strasbourg Cedex 2, France
}


\begin{abstract}
Surface spin waves in thin Permalloy films are studied by means of propagative spin wave spectroscopy. We observe a systematic difference of up to several tens of MHz when comparing the frequencies of counter-propagating waves. This frequency non-reciprocity effect is modeled using an analytical dipole-exchange theory that considers the mutual influence of non-reciprocal spin wave modal profiles and differences in magnetic anisotropies at the two film surfaces. At moderate film thickness (20 nm and below), the frequency non-reciprocity scales linearly with the wave vector and quadratically with the thickness, whereas a more complex non-monotonic behavior is observed at larger thickness. Our work suggests that surface wave frequency non-reciprocity can be used as an accurate spectroscopic probe of magnetic asymmetries in thin ferromagnetic films.
\end{abstract}

\maketitle


\section{Introduction}
The development of new types of microwave devices and logic elements based on propagating spin waves currently attract the efforts of many research groups~\cite{Schneider2008,Demidov2009,Klingler2014, Jamali,Sekiguchi2010, Khitun, Vlaminck2010}. One of the most peculiar properties of spin waves in thin films is the non-reciprocal propagation, which occurs when the spin wave wave vector $\mathbf{k}$ and the equilibrium magnetization $\mathbf{M}_\text{eq}$ lie in the film plane, perpendicular to each other. In this so-called magnetostatic surface wave (MSSW) configuration the amplitude and the modal profile of counter-propagating spin waves may differ considerably. This specific property might be an advantage for building nonreciprocal microwave devices such as isolators or circulators. Interestingly, the frequencies of the two counter-propagating spin waves also differ from each other as soon as the top/bottom symmetry of the ferromagnetic film is broken. This occurs when a metallic ground plane is brought in vicinity to one of the film surfaces~\cite{Amiri}, the saturation magnetization of the film is inhomogeneous~\cite{Kostylev2010}, the magnetic surface anisotropies at the two film surfaces are different~\cite{Hillebrands, Bailey}, or when an electrical current flows in the film (Oersted field effect)~\cite{Haidar2014}. While such frequency non-reciprocity has been predicted theoretically and observed experimentally in some very specific cases, a dedicated investigation of this phenomenon is still needed. Recently, this became particularly relevant with several reports on the extraction of the magnitude of the interfacial Dzyaloshinskii-Moria interaction (iDMI) from MSSW measurements in ultrathin films~\cite{di2015,belmeguenai2015,stashkevich2015,nembach2015}. Indeed, when included into the Landau-Lifshitz equation of motion, iDMI translates directly into a frequency non-reciprocity \cite{Moon,Kostylev2014}. However, this effect  always combines with the effect that we will discuss in the present paper, because both contributions to the spin wave frequency obey the same symmetry (odd in wave vector $\mathbf{k}$ and applied field $\mathbf H_{0}$). Therefore, it is of primary interest to determine precisely the magnitude of the frequency non-reciprocity induced by asymmetries of the magnetic properties across the film thickness, so as to be able to disentangle it from the iDMI non-reciprocity \cite{stashkevich2015}. In this article, we report systematic measurements of spin wave frequency non-reciprocity in thin Permalloy (Py) films. These measurements are interpreted with the help of a simple analytical theory accounting for different surface anisotropies at the top and bottom film surfaces.
\section{Experiment}
\begin{figure}[!b]
\includegraphics[width=8.5cm]{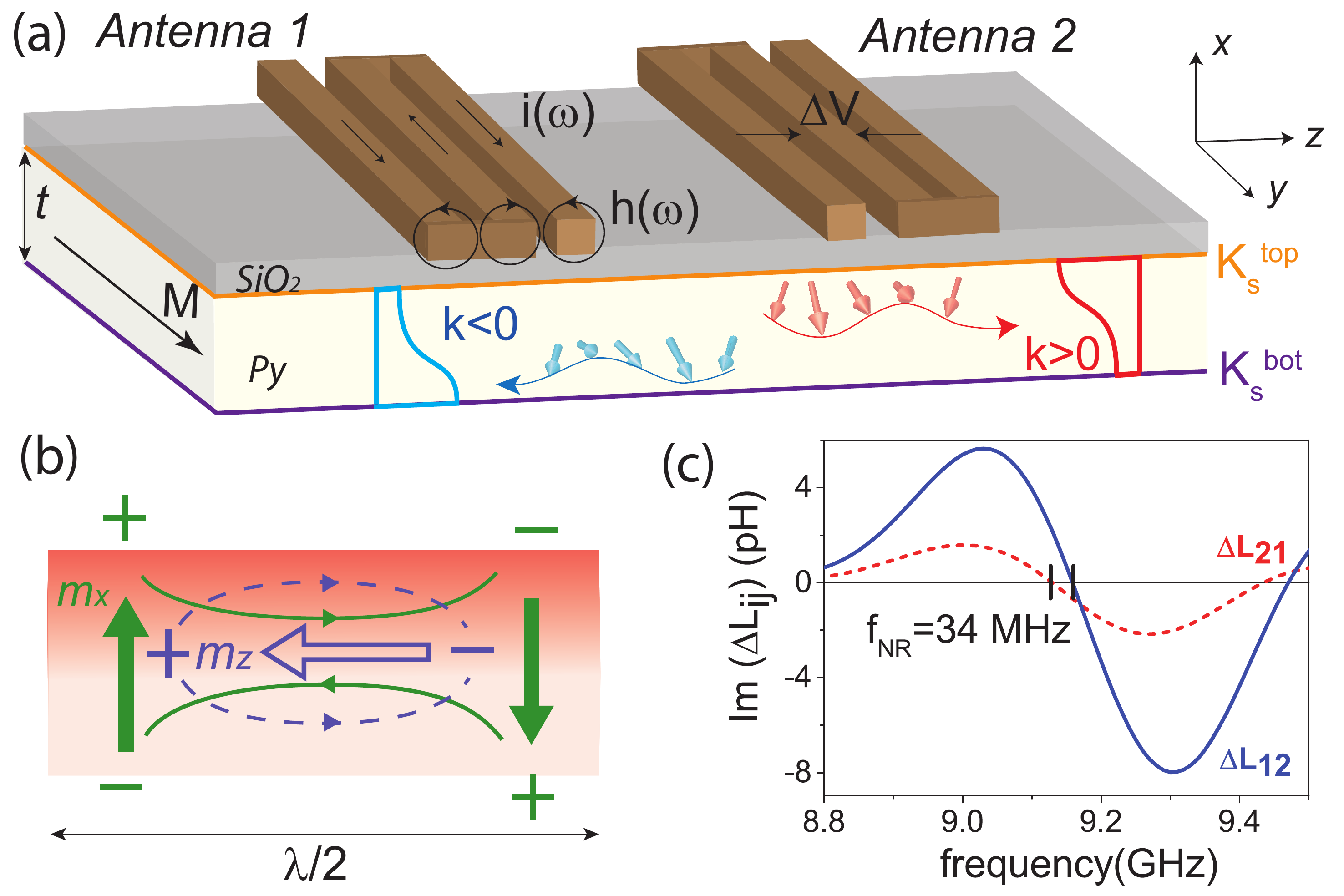}
\caption{(a) Schematic representation of the PSWS experiment in a Py strip: the spin wave is excited by a microwave current flowing in the emitting antenna. It propagates towards the receiving antenna. The asymmetric spin wave modal profiles across the film thickness are shown for $k>0$ and $k<0$ (b) Sketch of the dipolar field lines across half a wavelength for $k>0$ and $\mathbf{M}_\text{eq}$ along $+y$. Solid and open arrows indicate the directions of the $m_{x}$ and $m_{z}$ components of the dynamic magnetization. (c) Mutual-inductance spectra of a 40 nm thick Py strip for $k>0$ (solid line) and $k<0$ (dashed line) at $\mu_{0}H_0=37$ mT  and $|k|=3.86~\mu$m$^{-1}$.} \label{device}
\end{figure}

\begin{figure}
\includegraphics[width=8.5cm]{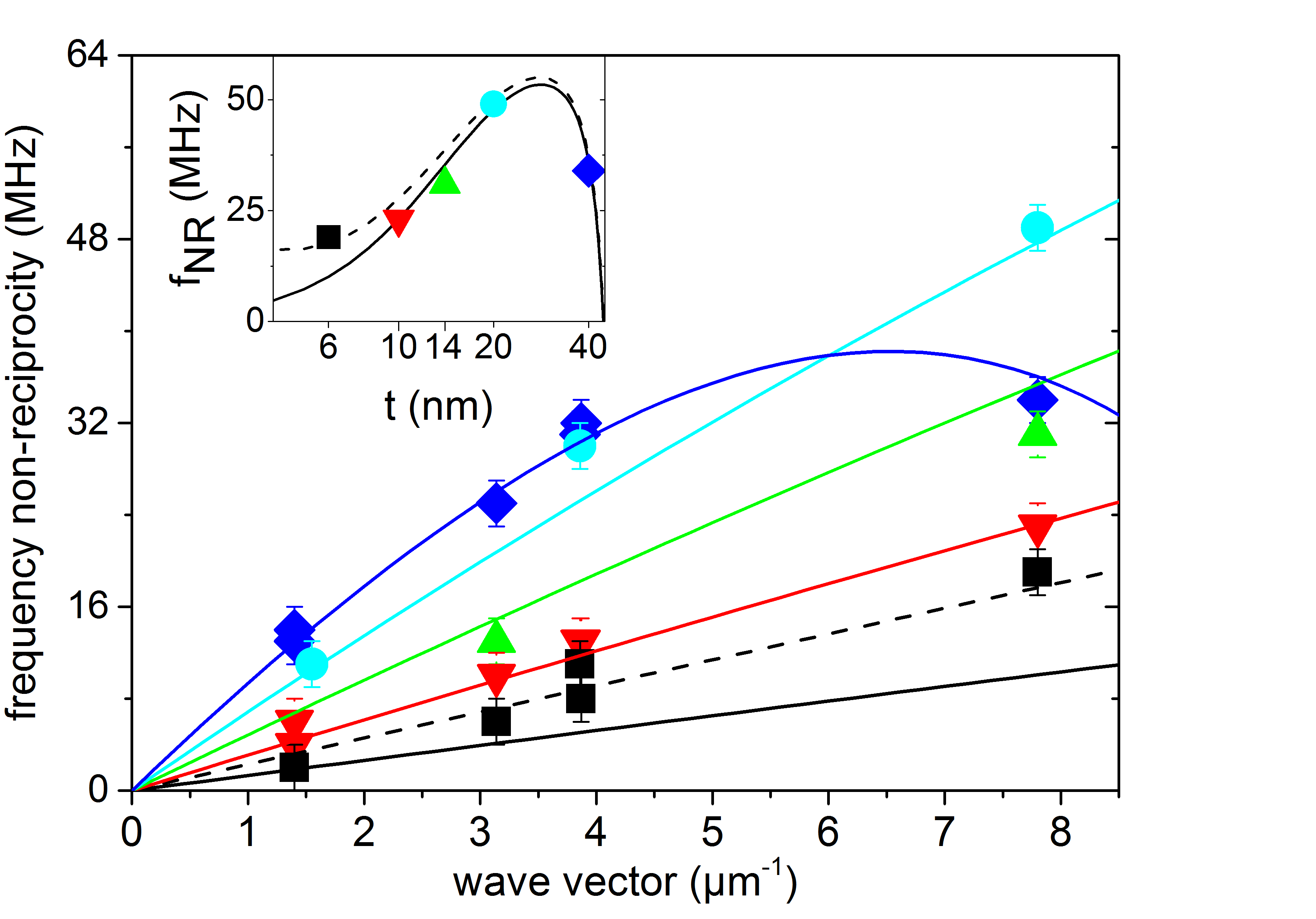}
\caption{Measured frequency non-reciprocity as a function of wave vector for $t=6$ nm (squares), $t=10$ nm (triangles down), $t=14$ nm (triangles up), $t=20$ nm (circles), and $t=40$ nm (diamonds). Solid lines show predictions of our model (Eq.~\ref{FN}) for $K_{\text s}^{\text {bot}}=0.15$ mJ/m$^2$, $K_{\text s}^{\text {top}}=-0.05$ mJ/m$^2$, $A=11.5$ pJ/m, $\mu_{0}M_{\text s}=1$ T, $\mu_{0}H_0=37$ mT, $\gamma/2\pi= 29.02$~GHz/T. (Inset) Same data plotted as a function of film thickness for $k=7.8~\mu$m$^{-1}$. The dashed lines are obtained when including iDMI ($D_{\text s}=-0.04$ mJ/m$^{2}$, see text for details), only for $t=6$~nm in the main panel.} \label{FN_fun_k}
\end{figure}

For this work, Al$_{2}$O$_{3}$(21~nm)/Py($t$)/Al$_{2}$O$_{3}$(5~nm) trilayers with $t=6-40$~nm have been sputter-deposited on intrinsic silicon substrates. Then, propagating spin wave spectroscopy devices consisting of a $2 - 8~\mu$m wide ferromagnetic strip and a pair of microwave antennas have been fabricated using standard lithography processes, as described in Refs. \onlinecite{Haidar2013,Haidar2014}.
The operational principle of our PSWS measurements is sketched in Fig.~\ref{device}(a). An external magnetic field $\mathbf{H}_{0}$ is applied across the strip in order to set the equilibrium magnetization $\mathbf{M}_\text{eq}$ in the MSSW geometry. From the microwave transmission between the two antennae, one extracts the mutual inductances $\Delta{L_{12}}$ and $\Delta{L_{21}}$, which correspond to a spin wave propagating with $k>0$ and $k<0$ respectively. In figure~\ref{device}(c), which shows typical spin wave spectra, one can clearly see that the wave propagating from antenna 1 to antenna 2 ($\Delta{L_{21}}$) is slightly shifted in frequency with respect to that propagating in the opposite direction ($\Delta{L_{12}}$). In this example, one measures a frequency non-reciprocity $f_\text{NR}=f_{12}-f_{21}=32\pm2$ MHz~\footnote{The value of the frequency non-reciprocity was determined using three different methods: visual inspection of the curves, calculation of their cross-correlation, and fits with sinusoidal functions under a gaussian envelope. The three methods give very similar results.}. We performed similar measurements for 12 devices with varying film thickness $t$ and wave vector $k=(1.5, 3.1, 3.9, 7.8)~\mu$m$^{-1}$. As in Refs.~\onlinecite{Vlaminck2010,Haidar2014}, we use data corresponding to both the main and the secondary peaks of the Fourier transform of the antenna geometry, and we use two different antenna pitches. This provides us with the four values of $k$. The results are reported as symbols in Fig.~\ref{FN_fun_k}. One observes a very strong dependence of $f_\text{NR}$ on wave vector and film thickness. For $t\leq 20$ nm, the frequency non-reciprocity increases linearly  with increasing $k$, whereas for $t=40$ nm, it first increases and then seems to saturate beyond $3.9~\mu$m$^{-1}$. $f_\text{NR}$ depends strongly on the film thickness too, as shown in the inset of Fig.~\ref{FN_fun_k}, where the data for $7.8~\mu$m$^{-1}$ are plotted as a function of $t$ on a semi-logarithmic scale. On the other hand, it depends neither on the stripe width (e.g. the data points for widths of 4 and 8 $\mu$m at $t=10$ nm and $k=3.9~\mu$m$^{-1}$ coincide exactly) nor on the magnitude of the applied field. As expected from the symmetry of MSSW non-reciprocity (see below), the effect systematically switches sign when $\mathbf{H}_0$ (and consequently $\mathbf{M}_\text{eq}$) is reversed (see Fig.~2 in Ref~\onlinecite{Haidar2014}). Thus, our main experimental finding is the observation of frequency non-reciprocities in the range of 3 to 49 MHz with a strong dependence on wave vector and film thickness. Before presenting the theoretical framework which will allow us to interpret these data, let us explain qualitatively the origin of the measured frequency non-reciprocity.

\section{Theory}
\subsection{Qualitative interpretation}
As was mentioned at the beginning, frequency non-reciprocity originates from the mutual influence of two factors: an intrinsic non-reciprocity of the modal profile and an asymmetry of the magnetic film properties. The origin of the modal profile non-reciprocity can be understood by examining the spatial distribution of the dipolar field generated by the dynamic magnetization. A sketch of how the oscillating magnetization is distributed across half a spin wave wavelength is shown in Fig.~\ref{device}(b) for $k>0$ and $\mathbf{M}_\text{eq}$ along $+y$ . Both $m_{x}$ and $m_{z}$ components of the dynamic magnetization (solid and open arrows, respectively) create magnetic poles which in turn generate a dynamic dipolar field (solid and dashed lines, respectively). In the upper half of the film the components of the dipolar field created by $m_{x}$ and $m_{z}$ add up, whereas they subtract in the lower half. As a result, for $k>0$, the dipolar field is larger in the upper half. For $k<0$, the situation is reversed: as $m$ rotates the other way round in the ($x$, $z$) plane,  the total dipole field is larger in the lower half of the film. This non-reciprocal asymmetry of the dynamic dipolar field is at the origin of the modal profile non-reciprocity: to build a true spin wave eigenmode, the dynamic magnetization has a tendency to compensate the dipole field asymmetry by increasing its amplitude on one side of the film (top or bottom depending on the sign of $k$). Note that the mode localization will be reversed if the equilibrium magnetization is switched  along $-y$.

Let us assume now that the magnetic properties are asymmetric across the thickness of the film, i.e. the dynamic and/or static effective fields are different in the top and bottom halves of the film cross-section. Quite naturally, the wave having larger amplitude in the half with higher effective field will oscillate at higher frequency than the wave having larger amplitude in the half with lower effective field, thus leading to the frequency non-reciprocity. This is illustrated in Fig.~\ref{device}(a) in the case of a homogeneous ferromagnetic film having out-of-plane uniaxial surface anisotropies of different magnitudes on both sides ($K_\text s^{\text {top}}$ and $K_\text s^{\text {bot}}$). Such a situation is expected to be quite common. Indeed, most  ferromagnetic film interfaces exhibit sizeable out-of-plane surface anisotropies, and their values are strongly dependent both on the composition of the adjacent non-magnetic layer and on the details of the interface structure (intermixing, roughness...).

\subsection{Quantitative interpretation}
For a quantitative estimate of the frequency non-reciprocity, we now resort to the theory of dipole-exchange spin waves~\cite{Kalinikos1980,Kalinikos1986}. We have recently revisited this theory to understand the MSSW modal non-reciprocity in thin ferromagnetic metal films~\cite{Kostylev2013} and to describe the non-reciprocal Oersted field induced frequency shift~\cite{Haidar2014}. In this paragraph, we will summarize the essential ingredients of this theory and introduce asymmetric surface anisotropies to derive an expression of the frequency non-reciprocity. The coordinate system used is shown in Fig.~\ref{device}(a): the $x$-axis is perpendicular to the film surface, the $y$-axis is along the applied magnetic field $\mathbf{H}_0$, and the $z$-axis coincides with the direction of propagation of the spin wave. To describe the magnetization dynamics in the film, we start with the linearized Landau-Lifshitz equation for plane spin waves of the form $\mathbf m=\mathbf m_{0}e^{i(\omega t-kz)}$:
\begin{equation}\label{LL}
i\omega\mathbf{m} = \gamma \mu_{0} H_{0}\mathbf{u}_{y} \times \mathbf{m}-\gamma \mu_{0} M_\text {s}\mathbf{u}_{y} \times \mathbf{h},
\end{equation}
where $\mathbf{m}$ and $\mathbf{h}$ are the dynamic components of the magnetization and effective field, respectively, $\gamma$ is the gyromagnetic ratio, $\mu_0$ is the permittivity of vacuum, and $M_\text s$ is the saturation magnetization. Note that vectors $\mathbf{m}$ and $\mathbf{h}$ have only two non-vanishing components $(m_{x},m_{z})$ and $(h_{x},h_{z})$, respectively.

The total dynamic field in Eq.~\ref{LL} may be written as
\begin{eqnarray}\label{h_tot}
&\mathbf{h}(x)&=\frac{2A}{\mu_{0} M_\text{s}^{2}} (\frac{\partial^{2}}{\partial x^{2}}-k^{2}) \mathbf{m}(x)+\int_{0}^{t} dx^{\prime} \overline{G}_{k}(x-x^{\prime})\mathbf{m}(x^{\prime}) \nonumber\\
&+&\frac{2}{\mu_{0}M_\text{s}^{2}}\mathbf{u}_{x}(K_{\text s}^{\text {bot}}\delta(x)m_{x}(0)+K_{\text s}^{\text {top}}\delta(x-t)m_{x}(t)),
\end{eqnarray}
where $A$ is the exchange stiffness constant and $\overline{G}_{k}$ is the magnetostatic Green's function. The first and second terms correspond to the exchange and dipolar fields respectively, and the last term describes the effective field generated by a uniaxial out-of-plane surface anisotropy at the top ($K_{\text s}^{\text {top}}$) and bottom ($K_{\text s}^{\text {bot}}$) surfaces \footnote{This effective field derives from an anisotropy energy $-K_{\text s} cos^{2} (\theta)$, where $\theta$ is the angle between the magnetization and the film normal. A positive $K_{\text s}$ corresponds therefore to an anisotropy with an easy axis along the film normal.}. Note the difference with the approach by Kalinikos and Slavin~\cite{Kalinikos1986}, where the surface anisotropies were treated as boundary conditions setting the exact form of the pinned exchange-modes forming the basis for the expansion. In the present case, we treat the surface anisotropy as an additional effective field affecting the unpinned exchange-modes. This perturbation approach simplifies the calculations a lot. It is well justified for a wide range of parameters, as we will see below.

Next, we expand the dynamic magnetization profile across the film thickness $\mathbf{m}(x)$ into a Fourier series forming an orthonormal basis of functions, and keep only terms up to first order
\begin{equation}\label{mag}
\mathbf{m}= m_{x}^{0}\mathbf{u}_{x}+m_{z}^{0}\mathbf{u}_{z}+ m_{x}^{1}\sqrt{2}\cos(\frac{\pi x}{t})\mathbf{u}_{x}+ m_{z}^{1}\sqrt{2}\cos(\frac{\pi x}{t})\mathbf{u}_{z}+...
\end{equation}
This expansion can be interpreted as a projection of $\mathbf{m}$ onto the first two unpinned exchange modes~\cite{Kalinikos1980,Kalinikos1986}: the $n=0$ FMR mode with a uniform profile and the so-called first perpendicular standing spin wave mode $n=1$ (PSSW1), which has an antisymmetric profile across the film thickness. The higher order terms in the Fourier series, which correspond to higher order PSSW modes ($n=2$ and above) can be safely neglected because of their much higher frequencies~\cite{Kostylev2013}.

Using the four terms of Eq.~\ref{mag} as a basis set, Eq.~\ref{LL} may be rewritten as a matrix eigenvalue equation, $i \Omega \bar{m} = \bar{\bar{C}} \bar{m}$ where $\Omega=\omega/(\gamma \mu_{0} M_\text{s})$ is the dimensionless frequency, $\bar{m}=(m_{x}^{0},m_{z}^{0},m_{x}^{1},m_{z}^{1})$, and  $\bar{\bar{C}}$ is the $4 \times 4$ dynamic matrix, which writes:
\begin{equation}\label{C}
\bar{\bar{C}}=
\begin{pmatrix}
\bar{\bar{C}}_{00}&\bar{\bar{C}}_{01}\\\bar{\bar{C}}_{10}&\bar{\bar{C}}_{11}
\end{pmatrix}
=
\begin{pmatrix}
0&\Omega_{z}^{0}&-iQ&0\\
-\Omega_{x}^{0}&0&-\delta&iQ\\
iQ&0&0&\Omega_{z}^{1}\\
-\delta&-iQ&-\Omega_{x}^{1}&0\\
\end{pmatrix},
\end{equation}
with
\begin{eqnarray}\label{omega}
&\Omega_{x}^{0}&=1-P_{00}-\varepsilon+h+\Lambda^{2}k^{2}, \nonumber\\
&\Omega_{z}^{0}&=P_{00}+h+\Lambda^{2}k^{2}, \nonumber\\
&\Omega_{x}^{1}&=1-P_{11}-2\varepsilon+h+\Lambda^{2}k^{2}+\frac{\Lambda^{2}\pi^{2}}{t^{2}}, \nonumber\\
&\Omega_{z}^{1}&=P_{11}+h+\Lambda^{2}k^{2}+\frac{\Lambda^{2}\pi^{2}}{t^{2}}. \nonumber
\end{eqnarray}
Here, $h=H/M_\text{s}$ is the dimensionless applied field, and $\Lambda=(2A/\mu_{0} M_\text{s}^{2})^{1/2}$ is the exchange length. The different terms in the $\bar{\bar{C}}$ matrix correspond to the projections of the different contributions to the effective field onto the four basis functions of Eq. \ref{mag}. More precisely, $P_{00}=1-\frac{1-e^{-kt}}{kt}$ and $P_{11}=\frac{(kt)^{2}}{\pi^{2}+(kt)^{2}}(1-\frac{2(kt)^{2}}{\pi^{2}+(kt)^{2}}\frac{1+e^{-kt}}{kt})$ are self-demagnetizing factors describing the average dipole field generated by the in-plane component of the uniform and PSSW1 modes, respectively. On the other hand, $Q=\frac{\sqrt{2}kt}{\pi^{2}+(kt)^{2}}(1+e^{-kt})$ is a mutual demagnetizing factor describing the dipolar interaction between  $n=0$ and $n=1$ basis functions. It corresponds precisely to the effect sketched in Fig.~\ref{device}(b), namely an antisymmetric in-plane dipole field component generated by a uniform out-of-plane magnetization component. The $\Lambda^{2} k^{2}$ and  $\Lambda^{2} \pi^{2}/t^{2}$ terms are the exchange contributions. Finally, $\varepsilon=\frac{2}{t\mu_{0}M_{\text s}^{2}}(K_{\text s}^{\text {bot}}+K_{\text s}^{\text {top}})$ and $\delta=\frac{2\sqrt{2}}{t\mu_{0}M_{\text s}^{2}}(K_{\text s}^{\text {bot}}-K_{\text s}^{\text {top}})$ are matrix elements related to the sum and the difference of the two surface anisotropies. The four $2 \times 2$ blocks composing the dynamic matrix have a direct interpretation: $\bar{\bar{C}}_{00}$ (resp. $\bar{\bar{C}}_{11}$)  is the dynamic matrix obtained by imposing a uniform (resp. PSSW1) profile for the magnetization across the film thickness. The corresponding eigenvalues $\Omega_{00}=\sqrt{\Omega_{x}^{0}\Omega_{z}^{0}}$ (resp.  $\Omega_{11}=\sqrt{\Omega_{x}^{1}\Omega_{z}^{1}}$)  are Kittel-like expressions of the mode frequencies in this approximation. The off-diagonal blocks $\bar{\bar{C}}_{01}=\bar{\bar{C}}_{10}^{\ast}$ describe the hybridization between the uniform and PSSW1 modes brought in both by the magnetic asymmetry $\delta$ and by the non-reciprocal dipole term $Q$. Solving for $det (\bar{\bar{C}}-i \Omega  \bar{\bar{1}})=0$, where $\bar{\bar{1}}$ is the identity matrix, one can write the dispersion relation for MSSWs in the following form:
\begin{equation}\label{disp}
(\Omega_{0}^{2}-\Omega^{2})(\Omega_{1}^{2}-\Omega^{2})+2\delta Q\Omega(\Omega_{z}^{1}-\Omega_{z}^{0})=0,
\end{equation}
where
\begin{align}\label{omega01}
&\Omega_{0,1}^2=\frac{\Omega_{00}^2+\Omega_{11}^2}{2}-Q^2\mp\nonumber\\
&\frac{1}{2}\sqrt{(\Omega_{11}^2-\Omega_{00}^2)^2-4Q^2((P_{00}-P_{11})^2+\frac{\Lambda^4\pi^4}{t^4})}
\end{align}
are the spin wave eigenfrequencies obtained for $\delta=0$. Considering the product $\delta Q$ as a small parameter, we get the following expression for the frequency shift induced by the magnetic asymmetry:
\begin{equation}\label{FN}
\Delta \Omega_{0}=\delta Q\frac{\Omega_{z}^{1}-\Omega_{z}^{0}}{\Omega_{1}^{2}-\Omega_{0}^{2}}.
\end{equation}
This expression constitutes the main theoretical finding of this paper. $Q$ being odd in $k$, this frequency shift changes sign when $k$ is reversed so that it corresponds effectively to a frequency non-reciprocity $\Omega_{NR}= \Omega_{0}(k<0)-\Omega_{0}(k>0)=-2 \Delta \Omega_{0}$. This frequency shift scales linearly with the difference in surface anisotropies. It is also inversely proportional to the difference between the two unperturbed spin wave frequencies $\Omega_{0}$ and $\Omega_{1}$, as expected from first order perturbation theory. As already noticed in the case of the Oersted field induced non-reciprocity~\cite{Haidar2013}, this frequency difference is governed by the exchange interaction, which tends to prevent hybridization. It is therefore essential to take exchange into account. An exchange-free theory would indeed predict unrealistic modal profiles, resulting in frequency non-reciprocities of incorrect amplitude and sometimes even of the wrong sign (see the discussion in Refs. \onlinecite{Haidar2014,Kostylev2013}). Finally, the numerator contains a factor $\Omega_z^{1}-\Omega_z^{0}$ which can be seen as an ellipticity factor associated with the fact that the surface anisotropy acts only on the $x$ component of the magnetization. In most cases, this factor remains positive. However, it changes sign close to the avoided crossing point~\cite{Kalinikos1986,Kostylev2013} in the dispersion relation $\Omega_{00}=\Omega_{11}$ and we find it to be responsible for the saturation observed for the 40 nm film at $k$ between 4 and 8 $\mu$m$^{-1}$. To verify these findings, we have calculated the spin wave  non-reciprocity  by discretizing the film thickness into 0.2 nm slabs and diagonalizing numerically the dynamic matrix corresponding to Eq. \ref{LL}. The frequency non-reciprocities thus determined are in good agreement with Eq.~\ref{FN} over most of the thickness range investigated. Deviations start to appear for thick films ($t=40$ nm) for which Eq. \ref{FN} underestimates the frequency non-reciprocity by a few tens of percents, probably due to the fact that the perturbation treatment of surface anisotropies becomes less valid. In the small thickness limit ($kt\ll 1$ and $\frac{\Lambda^{2}\pi^{2}}{t^{2}}\gg P_{00},P_{11},h,\varepsilon, \Lambda^{2}k^{2}$) one obtains from Eq.~\ref{FN}
\begin{equation}\label{FN-limit}
f_{\text {NR}}\simeq\frac{8\gamma }{\pi^{2}}\frac{K_{\text s}^{\text {bot}}-K_{\text s}^{\text {top}}}{M_{\text s}}\frac{k}{1+\frac{\Lambda^{2}\pi^{2}}{t^{2}}}.
\end{equation}
This asymptotic formula explains well the tendency observed in the thin film limit (Fig. \ref{FN_fun_k}): a linear dependence on the wave vector, and a nearly quadratic  dependence on the film thickness.

\begin{figure}
\includegraphics[width=8.5cm]{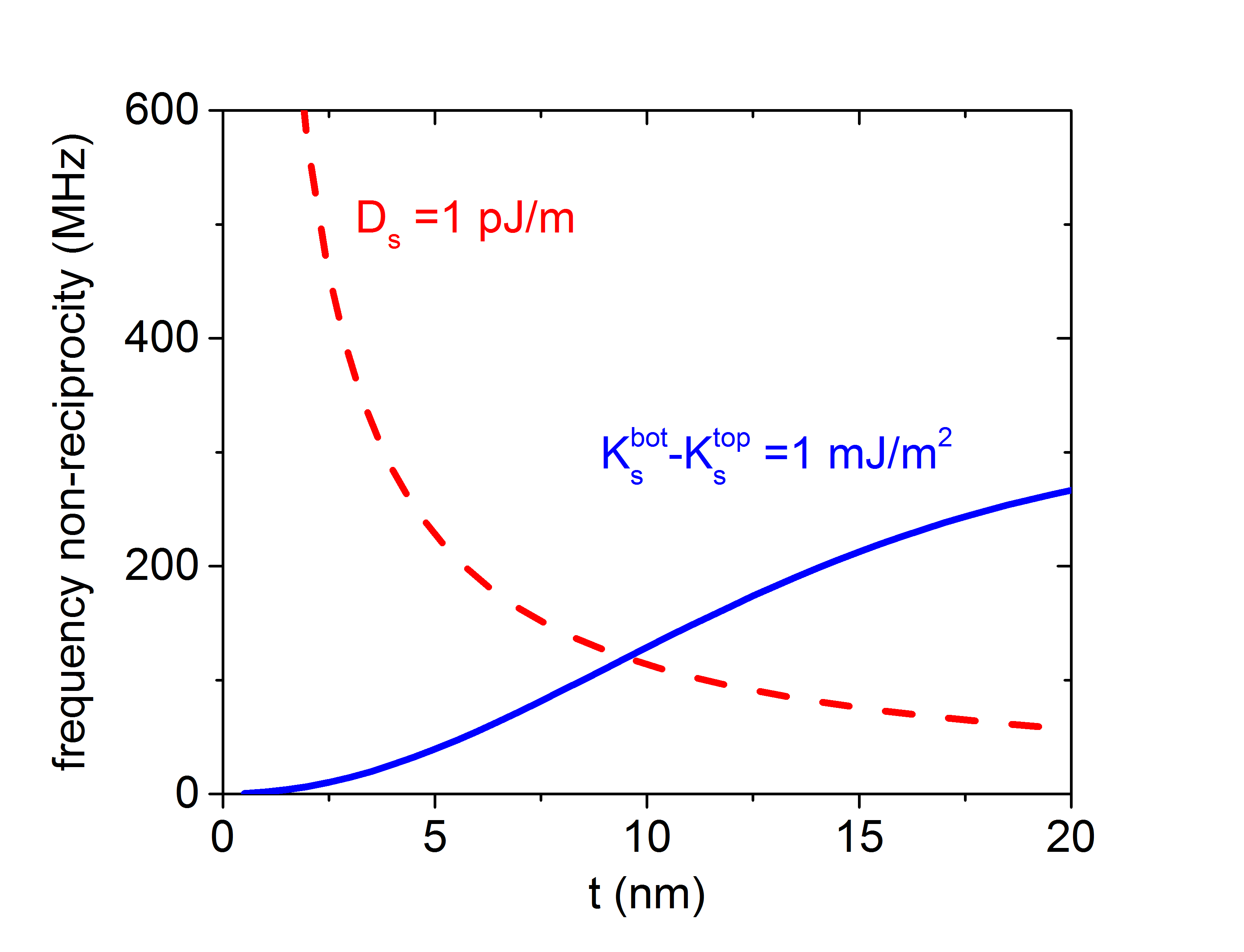}
\caption{Magnetic asymmetry and iDMI contributions to the frequency non-reciprocity for a typical ferromagnet/heavy metal bilayer with $A=11.5$~pJ/m, $\mu_0M_\text s=1$~T, $D_\text s=1$~pJ/m, $K_{\text s}^{\text {bot}}=1$~mJ/m$^2$, and $K_{\text s}^{\text {top}}=0$. The wave vector and applied magnetic field are $k=7.8~\mu$m$^{-1}$ and $\mu_0H_0=37$~mT, respectively. The magnetic asymmetry contribution (solid line) is calculated from Eq.~\ref{FN} whereas the iDMI contribution (dashed line) obeys the expression of $f_{\text {NR}}^{\text{DMI}}$ given in the text.} \label{FN_fun_t}
\end{figure}

Let us now discuss the quantitative comparison with the experimental results. As shown in Fig. \ref{FN_fun_k}, the model reproduces quite well the experimental data using a single value of the difference in surface anisotropies: $K_{\text s}^{\text {bot}}-K_{\text s}^{\text {top}}=0.20\pm0.01$ mJ/m$^2$.  The order of magnitude of this difference is consistent with typical values reported for surface anisotropies in permalloy films, which are in the range of $0-0.5$ mJ/m$^{2}$ (Refs.~\onlinecite{maranville,ingvarsson,hurdequint}). At first sight, the large difference may appear surprising for two nominally identical interfaces ($Py/Al_{2}O_{3}$ and  $Al_{2}O_{3}/Py$). However, surface anisotropies are known to depend quite strongly on the details of the interface structure, which in turn can dependent on the material deposition sequence. More specifically, we suspect a partial surface oxidation to play a role in our case \cite{Haidar2013}. Indeed, using  X-ray photoelectron spectroscopy and polarized neutron reflectivity, we have detected a non-magnetic nanometer-thick iron oxide forming mostly at the top surface, probably during the $Al_{2}O_{3}$ sputter deposition. In this picture, one expects a quite pronounced easy-axis surface anisotropy for the bottom $Al_{2}O_{3}/Py$ interface (as observed in most ferromagnet/non-magnetic oxide interfaces) and a reduced value for the top $Py/Al_{2}O_{3}$ interface. The sign of the observed frequency non-reciprocity is consistent with this expectation (i.e. $K_{\text s}^{\text {bot}}>K_{\text s}^{\text {top}}$). Note that the determination of the two individual surface anisotropies requires one to estimate their sum, in addition to their difference. This can be done by measuring  the effective magnetization of the films $M_{\text {eff}}=M_{\text s}-(K_{\text s}^{\text {bot}}+K_{\text s}^{\text {top}})/(\mu_0M_{\text s}t)$, using magnetometry techniques or ferromagnetic resonance. In the present case, we obtain $K_{\text s}^{\text {bot}}+K_{\text s}^{\text {top}}=0.1\pm0.1$ mJ/m$^2$,  the large error bar being due to the partial surface oxidation, which causes the average saturation magnetization to vary with film thickness.

Note the deviation between theory and experiment observed for $t=6$ nm (Fig.~\ref{FN_fun_k}). A possible explanation is the presence of a small Dzyaloshinskii-Moriya interaction. Indeed a value as small as $D_\text s=0.04$ pJ/m at one of the film interfaces generates an additional contribution (of the form given below) that explains the values measured experimentally (dotted lines in Fig.~\ref{FN_fun_k}). Because it scales as the inverse of the film thickness, this contribution becomes significant only for the thinnest film investigated. Such a small value of $D_\text s$, about fourty times smaller than the value observed in $Pt/Co/AlOx$ ultrathin films~\cite{belmeguenai2015}, seems plausible in a system which does not contain any heavy metal with strong spin-orbit interaction.

Let us finally comment on the implications of this work for recent determinations of the strength of the iDMI in ferromagnet/heavy metal systems based on MSSW frequency non-reciprocity measurements. We believe that asymmetric surface anisotropies are always present in such systems because the top and bottom interfaces systematically involve very different materials. The MSSW frequency non-reciprocity induced by iDMI is $f_{\text {NR}}^{\text{DMI}}=2 \gamma D_{\text s}k/(\pi M_{\text s}t)$ where $D_{\text s}$ is the micromagnetic iDMI constant expressed in J/m~\cite{belmeguenai2015}. It has the same wave vector dependence as the magnetic asymmetry contribution (Eq. \ref{FN-limit}), but a very distinct dependence on film thickness ($f_{\text {NR}}^{\text{DMI}}\propto t^{-1}$ versus $f_\text{NR} \propto t^{2}$ for the magnetic asymmetry contribution in the thin film limit). As a consequence, measurements conducted on ultrathin films (thickness of a few nm) will generally be dominated by the effect of iDMI, while measurements on moderately thin films (typically 20 nm) will generally be dominated by the effect of magnetic asymmetry, as suggested in Ref.~\onlinecite{stashkevich2015}. This is illustrated in Fig.~\ref{FN_fun_t}, which shows the magnitude of the two contributions calculated for typical values of $D_\text s$ and $K_{\text s}$ at ferromagnetic/heavy metal interfaces.

\section{Conclusion}
In conclusion, we have measured the frequency non-reciprocity of MSSWs as a function of film thickness and wave vector and we could account for the observed values using a simple analytical model of dipole-exchange spin waves in which asymmetric surface anisotropies are included as a perturbation. In the context of recent measurements of the iDMI interaction, we believe that the magnetic asymmetry contribution to MSSW frequency non-reciprocity is generally present and that it can be safely neglected only in the ultrathin film limit. From a more general point of view,  MSSW frequency non-reciprocity measurements on moderately thin films could be used to probe very accurately different kinds of magnetic asymmetries (differences of surface anisotropies but also gradients of saturation magnetization or magnetoelastic anisotropy). This spectroscopic method could nicely complement standard magnetic measurements which give access to quantities averaged over the film thickness.

\subsection{Acknowledgments}
This work was supported by the ANR (NanoSWITI, ANR-11-BS10-003) and the Australian Research Council. O. G. thanks IdeX Unistra for doctoral funding.

\newpage

\end{document}